 \documentclass[final,5p,times,twocolumn]{elsarticle}

\usepackage{amssymb}

\journal{Astroparticle Physics}

\begin{document}

\begin{frontmatter}



\title{Type IIn supernovae as sources of high energy astrophysical neutrinos}


\author{V.N.Zirakashvili, V.S.Ptuskin}

\address{Pushkov Institute of Terrestrial Magnetism, Ionosphere and Radiowave
Propagation, 142190, Troitsk, Moscow, Russia}

\begin{abstract}
It is shown that high-energy astrophysical neutrinos observed in the IceCube experiment
can be produced
 by protons accelerated in extragalactic Type IIn supernova remnants by shocks propagating
in the dense circumstellar medium.
 The nonlinear diffusive shock acceleration model is used for description of
particle acceleration. We calculate the neutrino spectrum produced by an
individual Type IIn supernova and
 the spectrum of neutrino background produced by IIn supernovae in the expanding Universe. We
also found that the arrival direction of one Icecube neutrino candidate
(track event 47) is at 1.35$^{\circ }$ from Type IIn supernova 2005bx.

\end{abstract}


\end{frontmatter}



\section{Introduction}

The detection of very high energy astrophysical neutrinos in the IceCube
experiment \cite{aartsen13, aartsen14,icecube15} opens up a new possibility for
investigation of particle acceleration processes in the Universe.
The neutrino production in cosmos is possible via the $pp$ and
$p\gamma$ interactions and the decay chains
$\pi^{+}\rightarrow \mu^{+} \nu_{\mu}$, $\mu^{+}\rightarrow e^{+}\nu_{e}\bar{\nu_{\mu}}$.
The observed astrophysical flux emerges from under more steep air show spectrum at about
$50$ TeV and has a cutoff at $2$ PeV. The neutrino typically carries a small part of the
primary proton energy, $E_{\nu}\approx 0.05E_{p}$, so the protons with energies up
to $E_{\max }\sim 10^{17}$ eV are required to explain the observations (this energy
is $\sim 10^{17}$ eV/nucleon in the case of neutrino production by nuclei).
Assuming an $E^{-2}$ power-law spectrum, the measured differential flux of
astrophysical neutrinos is $E^{2}F(E)=2.9\times 10^{-8}
\textrm{GeV}^{-2}\textrm{s}^{-1}\textrm{sr}^{-1}$
for the sum of the three evenly distributed neutrino flavors. The sources of observed neutrinos
are not yet identified. The detected $54$ events are scattered over the sky and do not show
any evident correlation with any astronomical objects \cite{icecube15, aartsen14b}.
It seems that Galactic
sources might account only for a minority of events. The detected astrophysical neutrinos
could be produced in extragalactic sources of ultra high energy protons and nuclei. The
discussion about potential sources of very high energy neutrinos in the light of the
last experimental results can be found in \cite{Anchor14,GaisserHalz14,murase14a}
where other useful references are given.

Rare extragalactic Type IIn supernova remnants are considered in the present paper as
 sources of diffuse high energy neutrinos. It is well established that supernova remnants
are efficient accelerators of protons, nuclei and electrons. They are the principle sources
of Galactic cosmic rays. The diffusive shock acceleration mechanism suggested
in \cite{krymsky77,bell78,axford77,blandford78} can provide the acceleration
of protons and nuclei in the most frequent Type IIP, Ia, Ib/c supernova events
up to about $10^{15}Z$ eV that allows to explain the spectrum and composition
of Galactic cosmic rays with a proton-helium knee at $3\times 10^{15}$ eV and
the maximum energy $\sim 10^{17}$ eV where iron nuclei dominate,
see \cite{ptuskin10}. Two orders of magnitude higher $E_{\max }\sim 10^{17}$ eV/nucleon
is needed to explain the IceCube neutrino observations. It can be achieved with the Type
IIn supernovae that stand out because of extremely dense wind of their progenitor stars
with a mass loss rate
$10^{-3} - 10^{-1}\ M_{\odot}$ yr$^{-1}$ \cite{moriya14}. As it will be shown below, the
large kinetic energy of explosion and very high gas density in the acceleration region
lead to the needed energy of accelerated particles and efficiency of neutrino production
in $pp$ interactions.

Diffusive shock acceleration  by supernova shocks propagating in
dense stellar winds was already considered in
\cite{murase11,katz11,murase14}. Simple analytical estimates
showed that radiowaves, gamma-rays and neutrinos  might be
observable from the nearest Type IIn supernova remnants if the
efficient diffusive shock acceleration takes place in these
objects. In the present paper we develop  this idea further and
investigate whether the flux of neutrinos produced in
extragalactic Type IIn supernova remnants can explain the IceCube
data. For this purpose we perform numerical modeling of particle
acceleration in a supernova remnant produced by Type IIn supernova
explosion and calculate neutrino production. Our model of
nonlinear diffusive shock acceleration describes the remnant
evolution and the production of energetic particles. The detailed
description of the model was presented in \cite{zirakashvili12}
and the simplified version of the model was used in
\cite{ptuskin10} for the explanation of energy spectrum and
composition of Galactic cosmic rays. Similar numerical models of
diffusive shock acceleration in supernova remnants were developed
and employed in \cite{berezhko94,kang06,berezhko07}.

The paper is organized as follows. In the next Sections 2 and 3 we
describe modeling of particle acceleration and calculate the
 spectrum of neutrinos produced in Type IIn supernova remnants.
These results are used in Section 4 for calculation of the diffuse
neutrino background in the expanding Universe.
The discussion of results and conclusions are given in Sections 5 and 6.

\section{Nonlinear diffusive shock acceleration model}

Details of our model of nonlinear diffusive shock acceleration can
be found in \cite{zirakashvili12}. The model contains coupled
spherically symmetric hydrodynamic equations  and the transport
equations for energetic protons, ions and electrons. The forward
and reverse shocks are included in the consideration.

The hydrodynamical equations for the gas density  $\rho (r,t)$, gas velocity $u(r,t)$,
gas pressure
$P_g(r,t)$, and the equation for isotropic part of the cosmic ray proton momentum distribution
 $N(r,t,p)$ in the spherically symmetrical  case are given by

\begin{equation}
\frac {\partial \rho }{\partial t}=-\frac {1}{r^2}\frac {\partial }{\partial r}r^2u\rho
\end{equation}

\begin{equation}
\frac {\partial u}{\partial t}=-u\frac {\partial u}{\partial r}-\frac {1}{\rho }
\left( \frac {\partial P_g}{\partial r}+\frac {\partial P_c}{\partial r}\right)
\end{equation}

\[
\frac {\partial P_g}{\partial t}+u\frac {\partial P_g}{\partial r}
+\frac {\gamma _gP_g}{r^2}\frac {\partial r^2u}{\partial r}=
\]
\begin{equation}
-(\gamma _g-1)\left( \Lambda (T)n^2+(w-u)\frac {\partial P_c}{\partial r}\right)
\end{equation}

\[
\frac {\partial N}{\partial t}=\frac {1}{r^2}\frac {\partial }{\partial r}r^2D(p,r,t)
\frac {\partial N}{\partial r}
-w\frac {\partial N}{\partial r}+\frac {\partial N}{\partial p}
\frac {p}{3r^2}\frac {\partial r^2w}{\partial r}
\]
\[
+\frac 1{p^2}\frac {\partial }{\partial p}p^3b(p)N+\frac {\eta _f\delta (p-p_{f})}{4\pi p^2_{f}m}\times
\]
\[
\rho (R_f+0,t)(\dot{R}_f-u(R_f+0,t))\delta (r-R_f(t))
\]
\begin{equation}
+\frac {\eta _b\delta (p-p_{b})}{4\pi p^2_{b}m}\rho (R_b-0,t)(u(R_b-0,t)-\dot{R}_b)\delta (r-R_b(t))
\end{equation}
Here $P_c=4\pi \int dpp^3vN/3$ is the cosmic ray pressure,
$w(r,t)$ is the advection velocity of cosmic rays, $T$, $\gamma
_g$ and $n$ are the gas temperature, adiabatic index and number
density respectively, $D(r,t,p)$ is the cosmic ray diffusion
coefficient. The radiative cooling of gas is described by the
cooling function $\Lambda (T)$.
 The function $b(p)$ describes the energy losses of particles.
In particular the energy losses due to
 $\it{pp}$ interactions and the radiative cooling are important at early
evolutionary stages of IIn supernovae.

Cosmic ray diffusion is determined by particle scattering on magnetic inhomogeneities.
The cosmic ray
streaming instability increases the
level of MHD turbulence in the shock vicinity \cite{bell78} and even significantly amplify the
absolute value of magnetic field in young supernova remnants \cite{bell04,zirakashvili08}. It
decreases the diffusion coefficient and increases the maximum energy of accelerated particles.
The results of continuing theoretical study of this effect can be found in review papers
\cite{bell2014,Caprioli2014}. In our calculations below, we use the Bohm value of the
diffusion coefficient $D_B=pvc/3qB$, where $q$ is the electric charge of particles.

Cosmic ray particles are scattered by moving waves and it is why the cosmic ray advection
velocity
 $w$ may differ from the gas velocity $u$ by the value of the radial
component of the Alfv\'en velocity
$V_{Ar}=V_A/\sqrt{3}$ calculated in the isotropic random magnetic field:
$w=u+\xi _AV_{Ar}$. The factor $\xi _A$
describes the possible deviation of the cosmic ray drift velocity from the gas velocity.
We  use values $\xi _A=1$ and $\xi _A=-1$ upstream of the
forward and reverse shocks respectively, where Alfv\'en waves are
generated by the cosmic ray streaming instability and propagate in
the corresponding directions. The damping of these waves heats the gas
upstream of the shocks \cite{mckenzie82} that is
described by the last term in Eq. (3).
The heating limits the total compression ratios of cosmic ray modified shocks.
In the downstream region of
the forward and reverse shock at $R_b<r<R_f$ we put $\xi _A=0$ and therefore $w=u$.

The magnetic field amplified by cosmic ray streaming instability upstream of the shock
is enhanced further via compression in the
 shock transition region. It can play a dynamical role downstream of the shock. We take
 magnetic pressure and magnetic energy flux into account downstream of the shock. This is
 a new element in comparison with our work \cite{zirakashvili12}
 where the magnetic field spatial distribution was prescribed.
The magnetic field is transported in the downstream region as the
gas with adiabatic index $\gamma _m$. Its impact  on the shock
dynamics is taken into account via the Hugoniot conditions.
Upstream of the forward shock where dynamical effects of
magnetic fields are small, the coordinate dependence of the
magnetic field $B$ can be described as:
\begin{equation}
B(r)=\sqrt{4\pi \rho _0}\frac {V_f}{M_A}\left( \frac {\rho (r)}{\rho _0}\right) ^{\gamma _m/2},
\end{equation}
Here $\rho _0$ and $\rho (r)$ are the undisturbed gas density at the shock position
and the density of the
medium where the shock propagates respectively, $V_f$ is the speed of the forward shock.
The parameter
  $M_A$ is similar to the Alfv\'en Mach number of the shock and
determines the value of the amplified magnetic field strength
far upstream of the shock. In the shock transition region the magnetic
field strength is increased by a  factor of $\sigma ^{\gamma _m/2}$,  where
$\sigma $ is the shock compression ratio. The expression similar
to Eq.(5) is also used in the upstream region of the reverse shock.

Below we
use the adiabatic index of isotropic random magnetic field $\gamma _m=4/3$.
For this value of the
adiabatic index, the magnetic pressure $P_m=B^2/24\pi $ is three times smaller
 than the  magnetic energy density ($=B^2/8\pi$).

Two last terms in Eq. (4)
correspond to the injection of thermal protons with momenta
$p=p_{f}$, $p=p_{b}$ and mass $m$ at the forward and
reverse shocks located at $r=R_f(t)$ and $r=R_b(t)$
respectively\footnote{We use indexes $f$ and $b$ for quantities
corresponding to the forward and
 reverse (backward) shock respectively.}. The
dimensionless parameters $\eta _f$ and $\eta _b$ determine the efficiency of injection.

We neglect the pressure of energetic electrons and treat them as test particles.
The evolution of the
electron distribution is described by equation analogous to Eq. (4)
with function $b(p)$  describing synchrotron and inverse Compton
(IC) losses and additional terms describing the production of secondary
leptons by energetic protons and nuclei.
The secondary electrons and positrons are effectively produced in the
dense medium of Type IIn supernova remnant via
 {\it pp} interactions. That is why we do not take into account injection of thermal
electrons at the shocks in the present calculations.

\section{Modeling of diffusive shock acceleration in the remnant of Type IIn supernova}

The blast wave produced by Type IIn supernova explosion propagates through the wind of the
presupernova star. We assume that the initial stellar wind density profile is described
by the following expression:

\begin{equation}
\rho =\frac {\dot{M}}{4\pi u_w\ r^2}.
\end{equation}
Here $\dot{M}$ is the mass-loss rate of the supernova progenitor star,
$u_w$ is the wind velocity and $r$ is the
 distance from the center of explosion.

We use the following parameters of the supernova explosion.
The explosion energy $E_{SN}=10^{52}$ erg,
 ejecta mass $M_{ej}=10\ M_{\odot}$, $\dot{M}=10^{-2}\ M_{\odot}$ yr$^{-1}$,
the parameter of ejecta
velocity distribution $k=9$ (this parameter describes the power-law density
profile $\rho_{s}\propto r^{-k}$ of the outer part of the ejecta that freely
expands after supernova explosion), the stellar wind velocity $u_w=100$ km s$^{-1}$
that are the characteristic values for Type IIn supernova \cite{moriya14}.

The injection efficiency is taken to be
independent of time $\eta _b=\eta _f=0.01$, and the particle injection
momenta are $p_{f}=2m(\dot{R}_f-u(R_f+0,t))$, $p_{b}=2m(u(R_b-0,t)-\dot{R}_b)$.
Protons of mass $m$ are
injected at the forward and reverse shocks.
The high injection efficiency
results in the significant shock modification already at early stage of the
supernova remnant expansion.

Figures (1)-(5) illustrate the results of our numerical calculations.
The value of the parameter $M_A=10$ was assumed.

The dependencies on time of the shock radii $R_f$ and $R_b$, the forward and
reverse shock velocities $V_f=\dot{R}_f$ and $V_b=\dot{R}_b$, cosmic ray energy
$E_{cr}/E_{SN}$ are shown in Fig.1.


Radial dependencies of physical quantities at  at
 $t=30$ yr are shown in Fig.2. The contact discontinuity between the ejecta and
the interstellar gas
is at $r=R_c=0.17$ pc. The reverse shock in the ejecta is located at $r=R_b=0.16$ pc.

We stop our calculations at $t=30$ yr when the stellar wind mass swept up
by the forward shock is of the
order of
 20 $M_\odot$. This value of the total mass loss is expected if the
initial mass of the Type IIn supernova progenitor
 is $70M_{\odot}$. The rest goes into ejecta ($\sim $ $10M_\odot $)
and the black hole ($\sim $ $40M_\odot $).
 At later times the forward shock enters rarefied medium
created by the fast tenuous wind of
 supernova progenitor at the main sequence stage.
So the production of neutrinos becomes negligible at this time.

\begin{figure}[t]
\includegraphics[width=8.0cm]{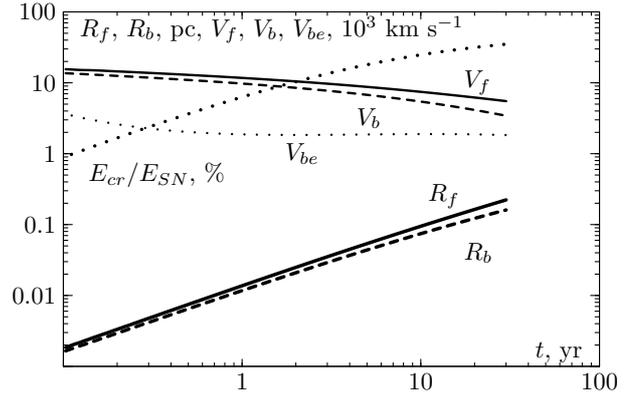}
\caption{Dependencies on time of the forward shock radius $R_f$ (thick solid line),
the reverse shock radius $R_b$ (thick dashed line), the forward shock velocity $V_f$
 (thin solid line), the reverse shock velocity $V_b$
 (thin dashed line) and the reverse shock velocity in the ejecta frame $V_{be}$
 (thin dotted line).
The ratio of cosmic ray energy and energy of supernova
explosion  $E_{cr}/E_{SN}$ (dotted line) is also shown.}
\end{figure}

\begin{figure}[t]
\includegraphics[width=8.0cm]{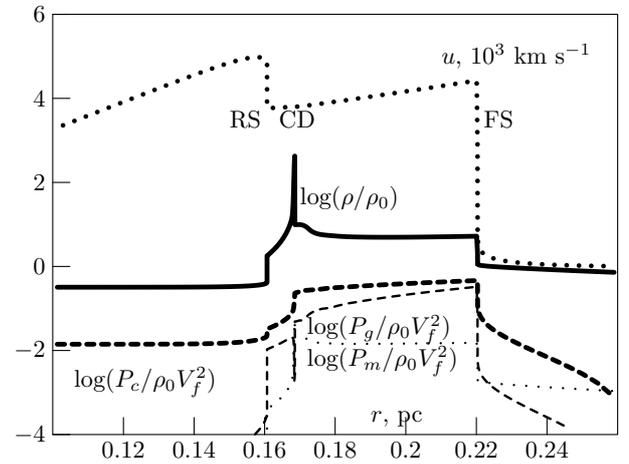}
\caption{Radial dependencies of the gas density (thick solid line), the gas
velocity (dotted line), cosmic ray pressure (thick dashed line) the magnetic
pressure (dotted line)
and the gas pressure (dashed line) at
 $t=30$ yr. At this point the forward shock velocity is $5.5\cdot 10^3$ km s$^{-1}$,
its radius is $0.22$ pc,
the magnetic field strength downstream of the forward shock is $0.06$ G,
the stellar wind density at the current
 forward shock position $\rho _0=10^{-20}$ g cm$^{-3}$.}
\end{figure}

\begin{figure}[t]
\includegraphics[width=8.0cm]{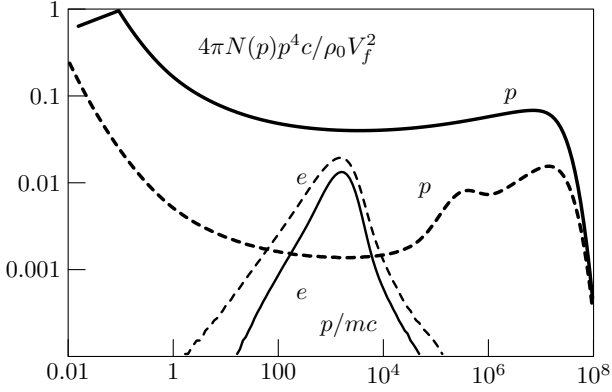}
\caption{Spectra of accelerated particles at $t=30$ yr.  The
spectrum of protons (thick solid line) and secondary electrons (multiplied
on $10^5$, thin solid line) at the forward shock, spectrum of protons (thick
dashed line) and spectrum of secondary electrons (multiplied on $10^5$, thin dashed
line) at the reverse shock are shown. }
\end{figure}

\begin{figure}[t]
\includegraphics[width=8.0cm]{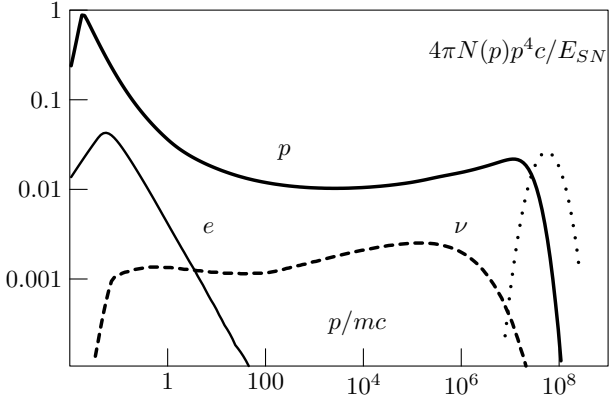}
\caption{ Spectra of particles produced in the supernova remnant
during $30$ yr after explosion. The spectrum of protons (thick
solid line ), the spectrum of secondary electrons
 (multiplied on $10^3$, thin solid line), the spectrum of neutrinos
(thick dashed line) are shown.}
\end{figure}

\begin{figure}[t]
\includegraphics[width=8.0cm]{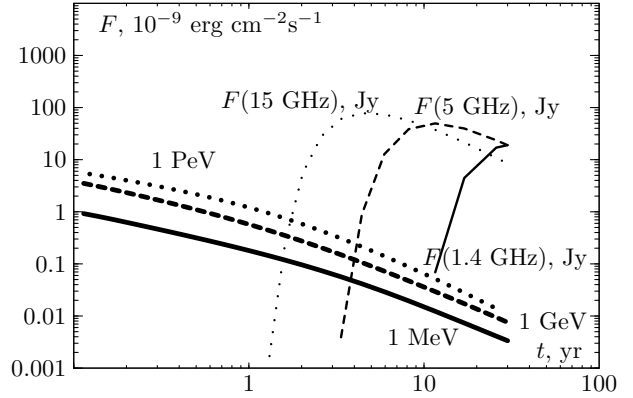}
\caption{Dependencies on time of fluxes from the supernova remnant at distance $1$ Mpc. We show
 the  neutrino flux at 1 PeV produced via pion decay (thick dotted line),
the gamma-ray flux at 1 GeV produced via pion decay
 (thick dashed line).
 The evolution of the synchrotron gamma-ray flux at $1$ MeV (thick solid line) and
 radio-fluxes  at $1.4$ GHz, $5$ GHz and $15$ GHz  (thin solid, dashed and dotted lines
respectively) are  also shown.}
\end{figure}

Spectra of accelerated protons and electrons at $t=30$ yr are shown in Fig.3.
At this point the maximum energy of accelerated protons is about $30$ PeV,
while higher energy particles have already left the remnant. The maximum energy of protons
accelerated
 at the reverse shock is close to $1$ PeV. However, the proton spectrum at the
reverse shock goes to higher
 energies because of the protons coming from the forward shock.

The spectra of particles produced during the life-time  of the remnant are shown in Fig.4.
They are calculated
as the sum of the spectra integrated throughout simulation domain and of the
time-integrated outward diffusive flux at the simulation
boundary at $r=2R_f$. About $25\%$ of the kinetic energy of explosion is transferred
to cosmic rays.

The temporal evolution of non-thermal emission produced in the supernova remnant at distance
$1$ Mpc is shown in Fig.5. We take into account a synchrotron
self-absorption and a free-free thermal absorption that are
 important for radio-supernovae \cite{chevalier98}. We use the temperature $T=10^4$K of
circumstellar
 wind that is a characteristic value for dense stellar winds ionized by radiation coming from
the forward and reverse shocks \cite{lundqvist88}.

It is instructive to compare the results of the present calculations
with the approximate analytical expression for the maximum energy
of protons accelerated by a supernova shock in the stellar wind.
Comparing the acceleration time
 $10D_B/V^2_f$ with the remnant age $t$ and with the time of $pp$ losses we found the
following expression for the maximum energy $E_{\max}$
\[
E_{\max }=\frac {3qV^2_f }{10M_Ac}\sqrt{\frac {\dot{M}}{u_w}}
\min \left( 1, \frac{t}{t_{pp}}\right) =
\]
\[
80\ \mathrm{PeV } \min \left( 1, \frac {t}{t_{pp}}\right)
\left( \frac {M_A}{10}\right) ^{-1}
\left( \frac {\dot{M}}{10^{-2}\ M_{\odot}\
\rm{yr}^{-1}}\right) ^{1/2}\times
\]
\begin{equation}
\left( \frac {u_w}{100\ \rm{km\ s}^{-1}}\right) ^{-1/2}
\left( \frac {E_{SN}}{10^{52}\ \rm{erg}}\right)
\left( \frac {M_{ej}}{10\ M_{\odot}}\right) ^{-1}
\end{equation}

where the time $t_{pp}$ is given by

\[
t_{pp}=\frac {0.5c\sigma _{pp}\dot{M}}{\pi u_wmV^2_f}=
0.2\ \mathrm{yr}
\left( \frac {\dot{M}}{10^{-2}\ M_{\odot}\
\rm{yr}^{-1}}\right) \times
\]
\begin{equation}
\left( \frac {u_w}{100\ \rm{km\ s}^{-1}}\right) ^{-1}
\left( \frac {E_{SN}}{10^{52}\ \rm{erg}}\right) ^{-1}
\left( \frac {M_{ej}}{10\ M_{\odot}}\right) .
\end{equation}

Here $\sigma _{pp}$ is the cross-section of {\it pp}
interactions. For simple estimates we  assume that the compression
 ratio of the forward shock equals to 4 and its velocity $V_f=\sqrt{2E_{SN}/M_{ej}}$
 is constant at the free expansion stage.


A similar estimate for Type Ia supernova explosion in the uniform medium gives
the "knee" energy $\sim 3$ PeV for protons. The shocks propagating in stellar
 winds accelerate particles to higher energies. In particular Type IIn
supernova remnants with their dense winds with mass-loss rate $\dot{M}
\sim 10^{-2}\ M_{\odot}$ yr$^{-1}$ can accelerate protons up to $80$
PeV according
 to Eq. (7). This order of magnitude estimate is in agreement
with our numerical results illustrated in Fig. 4. Thus, the explanation
of proton acceleration up to $\sim 10^{17}$ eV in Type IIn supernova
remnants is in line with the acceleration of cosmic rays up to the
knee in Type Ia supernova remnants.

\section{Calculation of background neutrino flux}

Using results of the previous Section, one can calculate the integrated
on time number of high-energy neutrinos emitted by a supernova
remnant that can be presented by the following equation:

\begin{equation}
Q(E_{\nu})=\int dt\int dE\int 4\pi r^{2}dr\frac{\rho (r,t)}{m}J_{cr}(E,r,t)\frac{d\sigma(E,E_{\nu})}{dE_{\nu}}.
\end{equation}

Here we introduced the cosmic ray flux distribution on energy $J_{cr}(E)=4\pi p^2 N(p)$ and
the effective cross section of neutrino production in $pp$ interactions
$\frac{d\sigma(E,E_{\nu})}{dE_{\nu}}$, $m$ is the proton mass.
The detailed description of the relevant cross sections and
kinematics of the neutrino production can be found in \cite{kelner}. The result of our
calculations is shown in Fig.4.

The neutrino spectrum produced by a single supernova remnant $Q(E_{\nu})$ can be
used for the determination of
extragalactic neutrino background. Distributed in the Universe
Type IIn supernova remnants give the
following diffuse flux of neutrinos:

\[
F(E_{\nu})=\frac {c}{4\pi H_0}\int ^{z_{\max }}_{0}dz \frac{Q((1+z)E_{\nu})\nu_{sn}(1+z)^{m}}
{\sqrt{\Omega_{m}(1+z)^{3}+\Omega_{\Lambda}}}
\]
\begin{equation}
=\frac{c}{4\pi H_0}\int^{(1+z_{\max })E_{\nu}}_{E_{\nu}}dE'\frac{E'^{m}}
{E_{\nu}^{m+1}}\frac{\nu_{sn}Q(E')}{\sqrt{\Omega_{m}E'^{3}/E_{\nu}^{3}+\Omega_{\Lambda}}}.
\end{equation}

Here the adiabatic energy loss of neutrinos produced at the redshifts
$0\leq z \leq z_{\max }$ is taken
into account. The present neutrino production rate per unit energy and
volume is $\nu_{sn}Q(E_{\nu})$,
where $\nu_{sn}$ is the rate of Type IIn supernovae at $z=0$ while the
cosmological evolution of the
sources in the comoving volume is described as
 $(1+z)^{m}$ ($m=0$ implies no evolution).
The evolution parameter $m=3.3$ for $z<1$ and no evolution at
$z>1$, the maximum redshift $z_{\max }=5$ and the rate $\nu
_{sn}=10^{-6}$ Mpc$^{-3}$ yr$^{-1}$ at $z=0$ are assumed in our
calculations (see e.g. \cite{dahlen12}). This rate of Type IIn
supernovae is $100$ times lower than the rate of all core collapse
supernovae. $H_{0}=70$ km s$^{-1}$ Mpc$^{-1}$ is the Hubble
parameter at the present epoch, the matter density in the flat
Universe is $\Omega_{m}=0.28$, and the $\Lambda$ -term is
$\Omega_{\Lambda}=0.72$.

The calculated background neutrino spectrum is shown in Fig.6.
The figure demonstrates a good fit
of our calculations to the IceCube data. We expect that the gamma-ray and neutrino
 background at energies below 100 TeV are produced by cosmic ray protons via $pp$
interactions in the interstellar medium of galaxies. This explains
why the first and the second IceCube data points are above our
theoretical curve. In addition the input of atmospheric neutrinos
can be significant at these energies.

\begin{figure}[t]
\includegraphics[width=8.0cm]{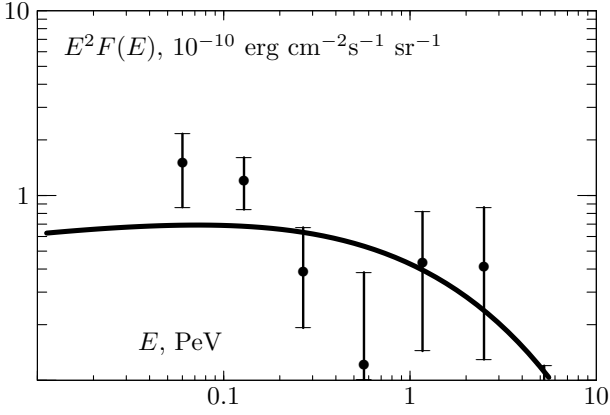}
\caption{ Calculated spectra of neutrino produced by IIn SNRs in
the expanding Universe (solid line).
IceCube 4 year data \cite{icecube15} are also shown. }
\end{figure}

Using the same approach we can calculate the flux of extragalactic protons.
Our results are compared
 with cosmic ray data in Fig.7. The proton flux produced by  extragalactic
IIn supernova is below
 the measured all particle cosmic ray flux and comparable to the measured proton
flux at energies
$2\cdot 10^{16}$ eV to $10^{17}$ eV. The calculated flux is not
corrected for possible magnetic horizon effect that can
considerably suppress the flux below about $10^{18}$ eV
\cite{aloisio05}. The suppression is due to strong deflection of
cosmic ray trajectories in extragalactic magnetic fields around
the sources in the expanding Universe.


With efficiency of cosmic ray production obtained in our calculations, Galactic Type
IIn supernovae
could make a significant contribution to the observed intensity of ultra high
energy cosmic rays.
However the intermittency of infrequent IIn supernova explosions
(one in 5 thousand years in the
Galaxy) makes the corresponding estimates rather uncertain.

Simple order of magnitude estimates can be done to clarify how the obtained neutrino
flux depends on supernova parameters. The main production of high energy particles
and neutrinos occurs up to the beginning of the Sedov stage when the
shock radius $R_{\mathrm{S}}$
can be determined from the condition
$M_{ej}=4\pi \int^{R_{S}} drr^{2}\rho=\dot{M}R_{\mathrm{S}}/u_{w}$. The time
 for the beginning of the Sedov stage $t_{\rm{S}}=R_{\rm{S}}/V_f$ can be written as
\[
t_{\mathrm{S}}=\frac {M_{ej}u_w}{\dot{M}V_f}=
10\ \mathrm{yr}
\left( \frac {\dot{M}}{10^{-2}\ M_{\odot}\
\rm{yr}^{-1}}\right) ^{-1}\times
\]
\begin{equation}
\left( \frac {u_w}{100\ \rm{km\ s}^{-1}}\right)
\left( \frac {E_{SN}}{10^{52}\ \rm{erg}}\right) ^{-1/2}
\left( \frac {M_{ej}}{10\ M_{\odot}}\right) ^{3/2}.
\end{equation}

We shall assume that at $t>t_{pp}$ the accelerated protons with the spectrum $E^{-2}$
are uniformly distributed in the  supernova shell.
The neutrino energy flux expected from a supernova at distance $D$ can be estimated as
(see also \cite{katz11,murase14})

\[
f(E_{\nu})E_{\nu}^2=\frac {3\xi _{CR}K_{\nu }}{8\pi \ln (E_{\max
}/mc^2)}\frac {V_f^3\dot{M}}{u_wD^2}
\left( 1+\frac{t}{t_{pp}}\right) ^{-1}=
\]
\[
10^{-8} \frac {\rm{erg}}{\rm{cm}^{2}\rm{s}}
\left( 1+\frac{t}{t_{pp}}\right) ^{-1}
D_{\rm{Mpc}}^{-2}\xi_{CR}
\left( \frac {\dot{M}}{10^{-2}\ M_{\odot}\ \rm{yr}^{-1}}\right) \times
\]
\begin{equation}
\left( \frac {u_w}{100\ \rm{km\ s}^{-1}}\right) ^{-1}
\left( \frac {E_{SN}}{10^{52}\ \rm{erg}}\right) ^{3/2}
\left( \frac {M_{ej}}{10\ M_{\odot}}\right) ^{-3/2}
\end{equation}
Here $\xi _{CR}$ is the ratio of cosmic ray pressure to the ram
pressure of the shock $\rho V_f^2$, $K_\nu \approx 0.25$ is the
fraction of energy that goes into  neutrinos in {\it pp}
interactions
and $E_{\max }$ is the maximum energy of accelerated protons given by Eq. (7).
The value of $\xi _{CR}$ is
$\xi _{CR}\sim 0.5$ in our numerical modeling of the efficient cosmic ray acceleration while
 a lower value $\xi _{CR}\sim 0.1$ is enough to explain the origin of
Galactic CRs in supernova remnants.

At early times $t<t_{pp}$ $pp$ losses dominate and the flux is almost steady.
It is interesting that the corresponding
luminosity $L_{\nu }\sim 10^{42}$ erg s$^{-1}$ is comparable with
the optical luminosity of Type IIn supernovae. This is not
surprising because
 both quantities are determined by the energetics of the forward shock.
The optical luminosity can be estimated from relation (see e.g. \cite{moriya14})

\begin{equation}
L= \epsilon \frac {\dot {M}V^3_{f}}{2u_w}.
\end{equation}
Here the factor $\epsilon \sim 0.1-0.5$.
This expression is often used to estimate the mass
loss $\dot{M}$ of Type IIn supernova progenitors.

Comparing with Eq. (12) we can rewrite the neutrino
 flux at $t<t_{pp}$ as

\[
f(E_{\nu})E_{\nu}^2=\frac {3\xi _{CR}K_{\nu }L}{4\pi D^2\epsilon\ln (E_{\max
}/mc^2)} =
\]
\begin{equation}
4\cdot 10^{-12} \frac {\rm{erg}}{\rm{cm}^{2}\rm{s}}\frac {\xi _{CR}}{\epsilon}
10^{0.4(13.7-m_{V})}
\end{equation}
Here we express the optical flux via the supernova visual magnitude 
 $m_V$. The similar expression can be used to estimate
unabsorbed fluxes of gamma-rays and synchrotron radiation produced
by secondary electrons and positrons.
The latter scans from radio band to GeV energies for Type IIn
supernovae. The unabsorbed gamma ray flux and the flux of
synchrotron radiation are lower by a factor of 1.5 and 6
respectively in comparison with the neutrino flux.

Integrated on time expression (12) can be used for the determination of neutrino
background produced by all Type IIn supernovae:

\[
F(E_{\nu})E_{\nu}^2=\frac {3\xi _{CR}K_{\nu }}{16\pi ^2\ln (E_{\max }/mc^2)}
\frac {\nu _{sn}c^2V_f\sigma _{pp}\dot{M}^2}{H_0mu^2_w} \times
\]
\[
\ln {\left( 1+\frac {t_{\mathrm{S}}}{t_{pp}}\right) }=
10^{-11}
\xi_{CR}\frac {\rm{erg}}{\rm{cm}^{2}\rm{s}\ \rm{sr}}
\ln {\left( 1+\frac {t_{\mathrm{S}}}{t_{pp}}\right) }
\times
\]
\[
\left( \frac {\dot{M}}{10^{-2}\ M_{\odot}\ \rm{yr}^{-1}}\right) ^{2}
\left( \frac {u_w}{100\ \rm{km\ s}^{-1}}\right) ^{-2}
\left( \frac {M_{ej}}{10\ M_{\odot}}\right) ^{-1/2}
\times
\]
\begin{equation}
\left( \frac {E_{SN}}{10^{52}\ \rm{erg}}\right) ^{1/2}
\left( \frac {\nu _{sn}}{10^{-6} \rm{Mpc}^{-3}\rm{yr}^{-1}}\right) ,
\end{equation}
For parameters used in our modeling the ratio $t_\mathrm{S}/t_{pp}\sim 50$. This
 is far from a so called calorimeter regime $t_\mathrm{S}/t_{pp}<1$ when a significant
 part of Type IIn supernova explosion energy is transferred to neutrinos.
The background  flux is several times higher if  the cosmological evolution of
supernova rate is taken into
account (see Eq. (10)).
These estimates are valid at neutrino energies $E_{\nu}<0.05E_{\max }$.
They confirm the efficiency
of high energy neutrino production in supernova explosion in the very dense
wind of a progenitor star,
as it takes place in the case of Type IIn supernova.

The rate of Type IIn supernovae $\nu _{sn}=10^{-6}$ Mpc$^{-3}$ yr$^{-1}$
 was adjusted in our calculations
to reproduce the IceCube observations.
Probably the real Type IIn supernova rate is several times higher.
However not all Type IIn supernovae have such high mass loss rates and
 explosion energies as we assumed.

\section{Discussion}

Our calculations show that the IceCube data can be explained by the
 neutrinos from Type IIn supernovae. If so, the arrival direction of every IceCube
neutrino coincides with the direction
 to some Type IIn supernova. However, we observe Type IIn supernovae
with the redshifts $z\leq 0.1$ while the background is determined by the
 supernovae with $z\sim 1$. Therefore we expect that only several percent
of observed neutrinos i.e. about one or two IceCube neutrino events are
associated with the known Type IIn supernovae. Unfortunately,
the arrival directions of majority of IceCube neutrinos are known with a low
 angular resolution (about 15$^{\circ }$).
Using ASIAGOSN supernova catalogue
\footnote{https://heasarc.gsfc.nasa.gov/W3Browse/all/asiagosn.html}
we found that several Type IIn supernovae
are indeed observed
in the direction of some IceCube neutrinos.
However this result has a low statistical significance.

Nevertheless we note that the arrival direction of one IceCube PeV neutrino candidate
(event 20) is
 within 5$^{\circ }$ from  SN 1978K. This  nearest Type IIn supernova in the galaxy
NGC 1313 at distance 4 Mpc is for
 decades observed in radio, X-rays and optics (e.g. \cite{smith07}). The mass-loss rate
of SN 1978K progenitor is
 estimated  as $\dot{M}=2\cdot 10^{-3}\ M_{\odot}$ yr$^{-1}$ \cite{chugai95} that is
5 times lower than we use in our
 calculations. Thus the fluxes expected from this supernova are not so high as ones shown
in Fig.5.
 In spite of this it is possible that the PeV neutrino candidate was emitted by SN 1978K.

The arrival direction of another IceCube PeV neutrino candidate (event 35) is
 within 10$^{\circ }$ from SN 1996cr. This  nearest Type IIn supernova in the
Circinus galaxy
 at the same distance 4 Mpc was also
 detected in radio, X-rays and optics (e.g. \cite{bauer08}). The Circinus galaxy is a
bright source
 of GeV gamma rays \cite{hayashida13}. It is known that the forward shock of
 SN1996cr entered the dense shell of circumstellar matter. The radio flux of this
supernova is only
 several times lower than the flux shown in Fig.5. That is why this IceCube PeV
neutrino candidate
 could be emitted by SN 1996cr.

We also looked for IceCube neutrinos in the vicinity of recent brightest Type IIn supernova
2010jl ($m_V=$13.5) and 2011fh ($m_V=$14.5).
There were 3 IceCube neutrinos (events 9, 11, 26)
within 25$^{\circ }$ from  2010jl supernova and 2 IceCube neutrinos (events 16, 49)
within 10$^{\circ }$ from supernova 2011fh. Using the measured
IceCube neutrino count rate 1.1 neutrino yr$^{-1}$ sr$^{-1}$ we estimate the expected number
 of 1.2 neutrino per 1.5 year in the vicinity of supernova 2010jl and 0.44 neutrino per
 4 year in the vicinity of supernova 2011fh. So it seems that there is some neutrino flux
enhancement in the direction of brightest Type IIn supernovae.
In addition the first neutrino (event 16)
 in the vicinity of supernova 2011fh was detected exactly at the time of the supernova discovery.

$14$ of $54$ IceCube neutrinos left tracks in the detector. The arrival directions
of these neutrinos
are determined with a good
 angular resolution (about 1$^{\circ }$). We found that an arrival direction
of one IceCube neutrino
candidate (track event 47) is at 1.35$^{\circ }$ from IIn supernovae 2005bx.
The mass-loss rate of this
 rather distant ($z\sim 0.03$) supernova was estimated as
$\dot{M}=0.037\ M_{\odot }$ yr$^{-1}$
while the wind velocity $u_w=813$ km s$^{-1}$ \cite{kiewe12}.
So the stellar wind density is
only  a factor
of 2  lower in comparison with one used in our calculations.
We estimate the radioflux at 5GHz
of the order of 1 mJy  using radiofluxes shown in Fig.5 and
 recalculating for 130 Mpc distance to supernova 2005bx.

The expected number of track events within 1.35$^{\circ }$ from known two hundred Type
IIn supernovae
 is 0.35. Taking into account only supernovae exploded between 2005 and 2014 we reduce
this number down to 0.25. The statistical significance will be higher
if during the future IceCube operation the neutrino emitted shortly (say 1 year)
after supernova explosion will be detected.


Association of the detected by IceCube doublet of neutrino track events  with
Type IIn supernova was also discussed in \cite{aartsen15}.

Type IIn supernova were not detected in Gev gamma-rays by Fermi LAT \cite{ackermann15}.
The upper limit derived corresponds to the ratio $\xi _{CR}/\epsilon \sim 1$ in Eq. (14).

\section{Conclusions}

Our main conclusions are the following:

1) The diffusive shock acceleration of particles in Type IIn supernova remnants
and the production
of neutrinos via $pp$ interactions in the dense presupernova winds can explain the
diffuse flux of high energy astrophysical neutrinos observed in the IceCube experiment.

2) The calculated maximum energy of protons accelerated in
  Type IIn supernova remnants is close to $10^{17}$ eV. This value is
higher than the maximum energy achieved in the main part of SNRs
and is explained by the high density of the circumstellar matter.

3) The efficient  acceleration of particles and production of secondary electrons and
positrons result in the fluxes
of radiowaves and gamma-rays that can be observed from the
nearest Type IIn supernova remnants.

4) Future IceCube operation and the search of correlations between neutrino arrival directions
 and the directions to Type IIn supernovae may check whether these objects are the
sources of high energy neutrinos but only one track event during $10$ years is expected
from these supernovae with the redshifts less than $0.1$. The bright phase of a
Type IIn supernova remnant as a source of PeV neutrinos lasts for about several years
after the supernova explosion.

The work was supported by the Russian Foundation for Basic Research grant 13-02-00056
and by the Russian Federation Ministry of Science and Education contract 14.518.11.7046.

\begin{figure}
\includegraphics[width=8cm]{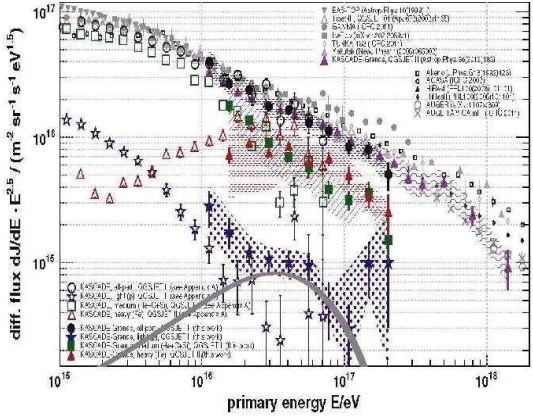}
\caption{ Comparison of the calculated cosmic ray proton background
produced by extragalactic Type
IIn supernovae
 (gray solid line) and data on cosmic ray protons and nuclei \cite{apel13}.}
\end{figure}












\end{document}